\documentclass[conference]{IEEEtran}
\IEEEoverridecommandlockouts

\usepackage{cite}
\usepackage{amsmath,amssymb,amsfonts}
\usepackage{algorithmic}
\usepackage{graphicx}
\usepackage{textcomp}
\usepackage{xcolor}
\def\BibTeX{{\rm B\kern-.05em{\sc i\kern-.025em b}\kern-.08em
    T\kern-.1667em\lower.7ex\hbox{E}\kern-.125emX}}
\usepackage{multirow}
\usepackage{makecell}
\usepackage{threeparttable}
\usepackage{hyperref}
\begin{document}

\title{Multichannel-to-Multichannel\\Target Sound Extraction \\Using Direction and Timestamp Clues 
\thanks{This work was supported by the National Research Foundation of Korea (NRF) grant (No. RS-2024-00337945) and STEAM research grant (No. RS-2024-00337945) funded by the Ministry of Science and ICT of Korea government (MSIT), the BK21 FOUR program through the NRF grant funded by the Ministry of Education of Korea government (MOE).}
}

\author{
\IEEEauthorblockN{Dayun Choi}
\IEEEauthorblockA{\textit{School of Electrical Engineering} \\
\textit{Korea Advanced Institute of Science and Technology} \\
Daejeon, Republic of Korea \\
cdy3773@kaist.ac.kr}
\and
\IEEEauthorblockN{Jung-Woo Choi}
\IEEEauthorblockA{\textit{School of Electrical Engineering} \\
\textit{Korea Advanced Institute of Science and Technology} \\
Daejeon, Republic of Korea \\
jwoo@kaist.ac.kr}
}

\maketitle

\begin{abstract}
We propose a multichannel-to-multichannel target sound extraction (M2M-TSE) framework for separating multichannel target signals from a multichannel mixture of sound sources. 
Target sound extraction (TSE) isolates a specific target signal using user-provided clues, typically focusing on single-channel extraction with class labels or temporal activation maps.
However, to preserve and utilize spatial information in multichannel audio signals, it is essential to extract multichannel signals of a target sound source. Moreover, the clue for extraction can also include spatial or temporal cues like direction-of-arrival (DoA) or timestamps of source activation.  
To address these challenges, we present an M2M framework that extracts a multichannel sound signal based on spatio-temporal clues. 

We demonstrate that our transformer-based architecture can successively accomplish the M2M-TSE task for multichannel signals synthesized from audio signals of diverse classes in different room environments. Furthermore, we show that the multichannel extraction task introduces sufficient inductive bias in the DNN, allowing it to directly handle DoA clues without utilizing hand-crafted spatial features. 

\end{abstract}


\section{Introduction}
Humans naturally focus on a specific sound in complex auditory environments with multiple sound sources. This ability allows us to attend to a target sound using clues like its time-frequency pattern or direction \cite{cocktail}. Target sound extraction (TSE) aims to mimic this by extracting the desired sound source using various types of clues. Common clues include class-labels \cite{uss, real-time}, a signal resembling the target \cite{soundbeam}, images or videos \cite{conceptbeam}, timestamps marking target occurrences \cite{timestamp1, timestamp2}, text descriptions \cite{describe, describe2}, directions or regions indicating the locations of targets \cite{rezero}, and combinations of these clues \cite{multimodal, clapsep, hetero, oct}.



However, these methods primarily focus on extracting a single-channel target signal. Multichannel mixtures, often recorded using microphone arrays, include spatial characteristics of a sound field. In applications like 3-D audio and virtual reality (VR) audio, interchannel relationships provide important spatial cues for rendering realistic sound. Similarly, in acoustic surveillance systems, interchannel time delays or phase differences are key to determining the direction or location of a target sound source. To fully exploit the spatial information in multichannel recordings, TSE should extract the multichannel source signal as if the microphone array had recorded the target sound alone.

There have been many DNN models designed to capture the spatial features and interchannel relations from multichannel sound sources, especially in speech separation or enhancement \cite{mcss1, mcss2, mcss3, mcss4, mcss5, mcss6, mcss7, mcss8} and DoA estimation \cite{mcdoa1, mcdoa2}.  
More recently, directional speech extraction \cite{rezero, dse} has also been proposed to extract speech from a multichannel mixture using its DoA clue. To incorporate DoA clue, its interchannel correlation features are extracted from a complex spectrogram of the mixture or individual one-hot embedding for each channel is integrated into spectral features.
Nevertheless, these models utilize interchannel information to extract a single-channel clean sound source or to determine the DoA of the sound source by removing reverberations and noises. 

Examples of separating a multichannel source signal can be found in binaural speech enhancement and TSE approaches \cite{semantic-hearing, binaural-se, binaural-filtering, interaural}. These approaches have demonstrated the potential of M2M extraction by preserving binaural cues. However, the binaural speech enhancement model \cite{binaural-se} focuses exclusively on the speech target, limiting its application to a specific source type. Binaural TSE \cite{interaural} overcomes this limitation and also utilizes spatial losses to minimize the degradation in binaural cues. Despite these advances, the primary clue for extraction remains the class-label related only to the time-frequency (TF) characteristics. In complex mixtures of signals from the same class, i.e., such as multiple instruments or speech sources, the direction of sound or timestamps of activation becomes compelling clues for extraction.  

\begin{figure*}
\centerline{\includegraphics[width=0.9\textwidth]{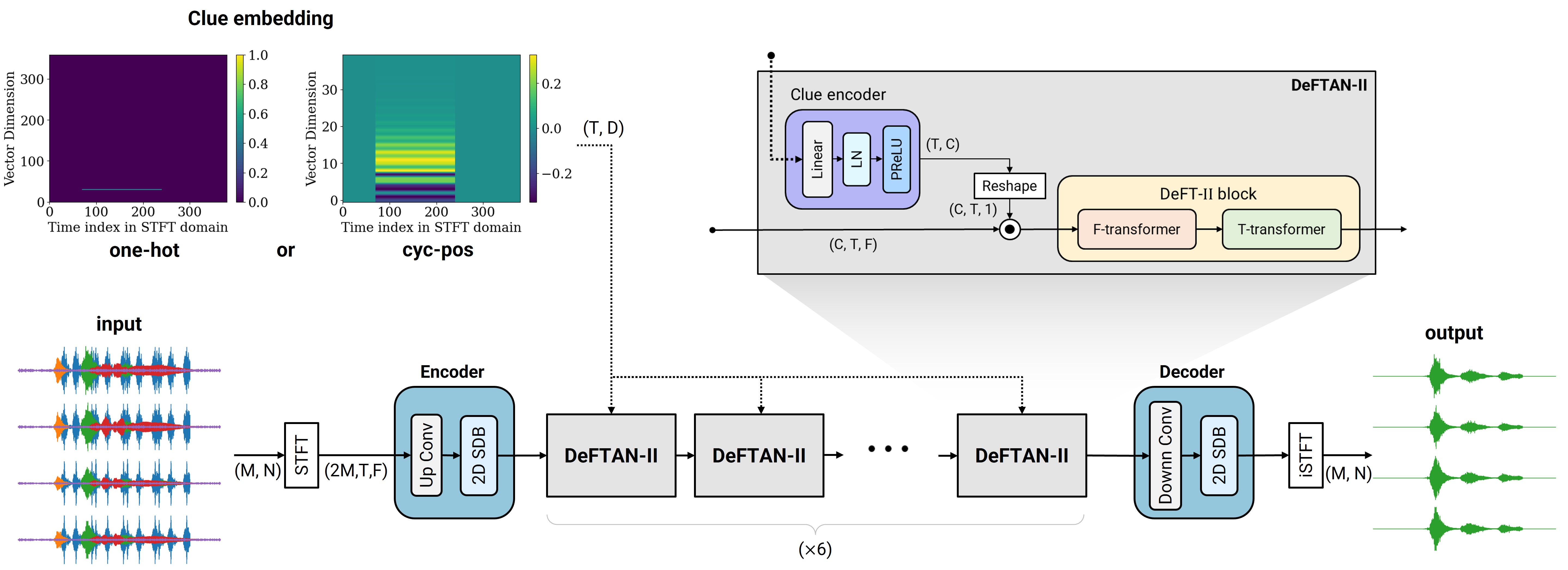}}
\caption{Model architecture for DoA-based multichannel-to-multichannel (M2M) target sound extraction.}
\label{fig_model}
\end{figure*}

To this end, we propose an M2M-TSE framework capable of extracting multichannel sounds from a complex, reverberant mixture using directional clues and timestamp information. We utilize dense frequency-time attentive network II (DeFTAN-II) \cite{mcss8} architecture as a backbone network, which delivers high performance with low time and space complexity in multichannel speech enhancement tasks. The key contribution of this work is the modified model designed to incorporate directional and timestamp clues for extracting multichannel sound sources of various types, rather than a single-channel speech. Furthermore, we demonstrate that the M2M extraction task encourages the model to capture and aggregate spatial features from the early layers of the architecture, enabling efficient extraction from simple directional and temporal embeddings without requiring additional spatial input features such as intensity vectors or interchannel correlation features.

\section{Proposed Methods}

\subsection{Problem Statement}
Our research focuses on extracting an $M$-channel reverberant sound signal $\mathbf{X}_i \in \mathbb{R}^{M \times N}$ of temporal length $N$, from an input mixture $\mathbf{Y} \in \mathbb{R}^{M \times N}$ of $I$ source signals captured from a microphone array in a reverberant room. The input mixture can be described as 
\begin{align} \label{eqn_mixture}
\begin{split}
\mathbf{Y} = \sum_{i=1}^I \mathbf{X}_i + \mathbf{V}
= \sum_{i=1}^I \mathbf{s}_i*\mathbf{R}_i + \mathbf{V},
\end{split}
\end{align}
where $\mathbf{s}_i \in \mathbb{R}^{N}$ and $\mathbf{R}_i \in \mathbb{R}^{M \times N}$ are the $i$-th dry source signal and its $M$-channel room impulse response (RIR), respectively. Here, $\mathbf{V} \in \mathbb{R}^{M \times N}$ is the multichannel measurement noise, and $\ast$ indicates the temporal convolution. 
When the target sound is the $g$-th source $\mathbf{X}_{g}$ (${g} \in \{1,\cdots,I\} $), 
the multichannel signal $\hat{\mathbf{X}}_{g} \in \mathbb{R}^{M \times N}$ extracted from a TSE model with model parameters $\theta$ can be written as 
\begin{align} \label{eqn_TSE}
\hat{\mathbf{X}}_{g} = \textup{TSE}(\mathbf{Y}, \mathbf{C}_{g}; \theta),
\end{align}
where $\mathbf{C}_{g}$ is a clue embedding derived from DoA and temporal activity (timestamps) of the target source. In this work, we consider the clue embedding $\mathbf{C}_{g}$ given by either a one-hot or a cyclic positional vector, as described in section \ref{subsec_clues}.


\subsection{Model Architecture} \label{subsec_model}
The backbone of the proposed network is the DeFTAN-II \cite{mcss8} architecture. DeFTAN-II is a transformer-based architecture that performs complex spectral mapping to extract a single-channel clean speech with suppressed noise and reverberation. Our TSE objective is similar, but the main difference is that a multichannel target signal should be extracted and a clue for identifying the target signal should be injected into the network. To achieve these, we introduce the following modifications to the backbone network.  

The overview of the proposed architecture is depicted in Fig.~\ref{fig_model} without the batch dimension. First, a multichannel input waveform is converted into a complex spectrogram of dimensions $2 M \times T \times F $ by short-time Fourier transform (STFT), where $T$ and $F$ denote the number of time and frequency bins, respectively. This spectrogram is then transformed to a tensor with increased channel dimension ($C$) using a 2-dimensional split dense block (2D SDB) encoder. 2D SDB is a modified version of DenseNet \cite{dense} introduced for extracting spatial features and learning local spectral-temporal relations. Since the channel dimension includes encoded spatial features and local time-frequency (TF) information, we combine the clue embedding with the channel dimension of the encoded tensor to extract the target sound of a specific direction.

In a series of DeFTAN-II blocks, the F- and T-transformers analyze the relationships in spectral and temporal sequences.  
We repeatedly embed the clue during this stage, to gradually align the features developed by DeFTAN-II blocks with those of the target sound. Finally, the aligned features are decoded to a multichannel waveform through a decoder reducing the channel dimension from $C$ to $2M$ as in the original STFT, and the inverse STFT (iSTFT) operation.

\begin{figure}[htbp]
\centerline{\includegraphics[width=\columnwidth]{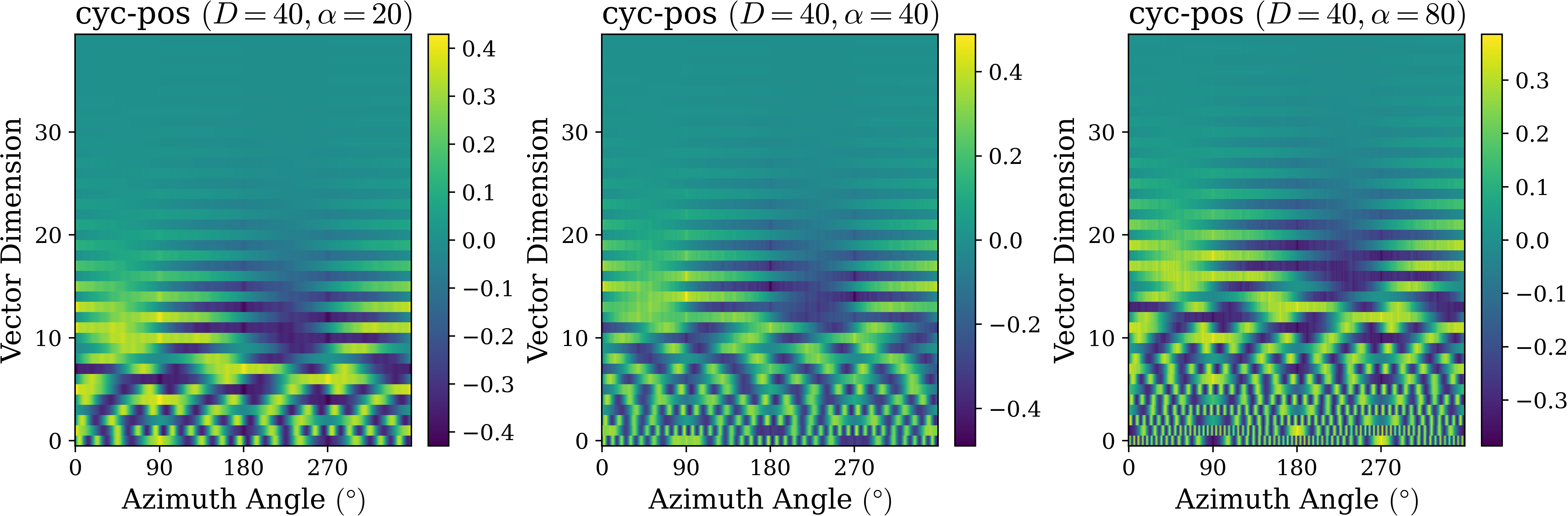}}
\caption{
Clue embeddings given by 
cyclic positional embedding vectors for different 
scaling factors ($\alpha$).
}
\label{fig_clues}
\end{figure}

\subsection{Spatio-temporal Clues} \label{subsec_clues}
The clue utilized in this work is the direction and timestamps of a target source. For simplicity, we consider only azimuth angles as the direction clue and encode them in two different ways: one-hot or cyclic positional encoding.   
For the one-hot encoding, the clue embedding vector $\mathbf{1}_{\textup{one-hot}}(\phi) \in \mathbb{R}^{360}$ defined for direction $\phi \in [0, 360)$ with $1^{\circ}$ resolution has a value of one only in the index corresponding to the direction $\phi$, with all other values being zero. That is, for the index $j \in [0, 360)$, 
\begin{align} \label{eqn_one-hot}
\mathbf{1}_{\textup{one-hot}}(\phi, j) = 
\left\{\begin{matrix}
1 \; &\textup{where} \; j=\phi, \\ 
0 \; &\textup{otherwise}.
\end{matrix}\right.
\end{align}

While the one-hot vector contains unique information for each direction, it cannot represent periodicity correctly, showing the abrupt transition from $359^{\circ}$ to $0^{\circ}$. To address this problem, we employ the cyclic positional (cyc-pos) encoding \cite{cyc-pos}. The cyc-pos vector $\mathbf{PE}_{\textup{cyc-pos}}(\phi) \in \mathbb{R}^{D}$ for embedding dimension $D$ can be represented as:
\begin{align} \label{eqn_cyc-pos}
\begin{split}
\mathbf{PE}_{\textup{cyc-pos}}(\phi, 2j) &= \sin(\sin(\phi) \cdot \frac{\alpha}{10000^{2j/D}}), \\
\mathbf{PE}_{\textup{cyc-pos}}(\phi, 2j+1) &= \sin(\cos(\phi) \cdot \frac{\alpha}{10000^{2j/D}}),
\end{split}
\end{align}
where $j \in [0, \frac{D}{2})$, and $\alpha$ is the scaling factor controlling the angular range utilized for the positional encoding. In Fig.~\ref{fig_clues}, examples of cyclic positional encoding are presented across various $\alpha$. Both $D$ and $\alpha$ are hyper-parameters determined empirically, and in this work, $D=40$ and $\alpha=20$ were selected from the parameter study. The generated positional embedding is normalized by its L2 norm for each direction.  

To further reduce ambiguity in signal extraction, the encoded positional embedding vector is combined with the timestamp clue. The timestamp indicates the occurrences of a target signal, so the positional embedding is broadcasted along the time dimension, such that rows of the final embedding matrix $\mathbf{C}_{g} \in \mathbb{R}^{T \times D}$ are nonzero only when the target source is active. One example of the embedding matrix is shown in the top left corner of Fig.~\ref{fig_model}.  
This embedding is encoded by linear layers, followed by layer normalization (LN) \cite{layernorm} and the parametric rectified linear unit (PReLU) \cite{prelu} activation, and is then multiplied element-wise with the output of the encoder and DeFTAN-II blocks except for the final block, across the channel and the time dimensions.

\section{Experiment and Analysis}

\subsection{Datasets}\label{FAT}
Target signals of training and test datasets were collected from the FSD Kaggle 2018 \cite{fsd-kaggle} dataset, a set of sound sources with 41 classes. Following a setup similar to that used for generating reverberant speech in \cite{spatialize}, a 4-channel circular microphone array with a radius of 10 cm was positioned in cuboid rooms of which width, depth, and height dimensions were randomly sampled from the uniform distribution within the ranges [5, 10] m, [5, 10] m, and [3, 4] m, respectively. The distance between the center of the microphone array and all sources was also randomly varied within the range of [0.75, 2.5] m, and the minimum angle between two sources relative to the array was set to $20^\circ$. The RIRs were generated using the image source method implemented in the \verb|pyroomacoustics|\footnote{\url{https://github.com/LCAV/pyroomacoustics}} \cite{pyroom} library, and the reverberation time (RT60) of each room was varied between [0.2, 1.3] s. All sound sources convolved with RIRs were mixed using the \verb|scaper|\footnote{\url{https://github.com/justinsalamon/scaper}} \cite{scaper} library depending on their timestamp, duration, and magnitude in dB. In addition, noise signals obtained from the 1st, 3rd, 5th, and 7th microphone noises in the REVERB challenge \cite{reverb} dataset were added. All input mixtures were 6-second-long samples, sampled at 8 kHz for fast computation. The numbers of mixtures constituting the training, validation, and test datasets were 12.5K, 5K, and 2.5K, respectively.


\begin{table*}[htbp]
\begin{threeparttable}
\caption{Performance comparisons across TSE models and clues.}
\vspace{-0.5cm}
\setlength{\tabcolsep}{7.pt}
\renewcommand{\arraystretch}{1.2}
\centering
\begin{tabular}{c|c|cc|cc|cccc}
\hline
\multirow{2}{*}{Model} & \multirow{2}{*}{\makecell{Type of  Clue}} & \multicolumn{2}{c|}{Parameters} & \multicolumn{2}{c|}{SNR Metrics $\uparrow$} & \multicolumn{4}{c}{Spatial Errors $\downarrow$} \\ \cline{3-10} 
                       & & \multicolumn{1}{c|}{$D$} & $\alpha$ & \multicolumn{1}{c}{SNRi (dB)} & SI-SNRi (dB) & \multicolumn{1}{c}{$\Delta$ILD (dB)} & \multicolumn{1}{c}{$\Delta$IPD (rad)} & \multicolumn{1}{c}{$\Delta$ITD ($\mu$s)} & $\Delta$ITD-GCC ($\mu$s)  \\ 
                       \hline\hline
 & TS only & \multicolumn{1}{c|}{40} & - & \multicolumn{1}{c}{9.98} & 6.65 & \multicolumn{1}{c}{0.95} & \multicolumn{1}{c}{0.90} & \multicolumn{1}{c}{239.46} & 234.23 \\ 
                            Waveformer\cite{semantic-hearing} & one-hot + TS & \multicolumn{1}{c|}{360} & - & \multicolumn{1}{c}{11.35} & 7.90 & \multicolumn{1}{c}{0.81} & \multicolumn{1}{c}{0.89} & \multicolumn{1}{c}{175.68} & 185.13 \\ 
                            & cyc-pos + TS & \multicolumn{1}{c|}{40} & 20 & \multicolumn{1}{c}{11.90} & 8.73 & \multicolumn{1}{c}{0.73} & \multicolumn{1}{c}{0.87} & \multicolumn{1}{c}{161.33} & 190.48 \\ 
                            \hline 
 & TS only & \multicolumn{1}{c|}{40} & - & \multicolumn{1}{c}{16.85} & 14.28 & \multicolumn{1}{c}{0.42} & \multicolumn{1}{c}{\textbf{0.77}} & \multicolumn{1}{c}{105.91} & 122.76 \\ 
                            & one-hot + TS & \multicolumn{1}{c|}{360} & - & \multicolumn{1}{c}{12.23} & 8.45 & \multicolumn{1}{c}{0.91} & \multicolumn{1}{c}{\underline{0.84}} & \multicolumn{1}{c}{164.93} & 154.81 \\ \cline{2-10}
                            \makecell{Proposed\\} &  & \multicolumn{1}{c|}{} & 80 & \multicolumn{1}{c}{14.81} & 12.96 & \multicolumn{1}{c}{0.79} & \multicolumn{1}{c}{0.91} & \multicolumn{1}{c}{96.78} & 108.32 \\ 
                            & cyc-pos + TS & \multicolumn{1}{c|}{40} & 40  & \multicolumn{1}{c}{\underline{17.22}} & \underline{15.85} & \multicolumn{1}{c}{\underline{0.37}} & \multicolumn{1}{c}{0.88} & \multicolumn{1}{c}{\underline{80.97}} & \textbf{103.77} \\ 
                            & & \multicolumn{1}{c|}{} & 20  & \multicolumn{1}{c}{\textbf{17.78}} & \textbf{16.51} & \multicolumn{1}{c}{\textbf{0.32}} & \multicolumn{1}{c}{0.87} & \multicolumn{1}{c}{\textbf{77.37}} & \underline{106.63} \\ 
                            \hline
\end{tabular}
\noindent{Boldface and underlined numbers indicate the best and the second-best results for each metric, respectively. TS denotes the timestamp clue.  
}
\label{tab_result}
\end{threeparttable}
\end{table*}

\subsection{Implementation Details}
All experiments were conducted in PyTorch framework using automatic mixed precision (AMP) training on a GeForce RTX 4090. The training parameters included a batch size of 4, the Adam optimizer with an initial learning rate of 0.0005, multiplied with 0.1 when the scale-invariant signal-to-noise ratio (SI-SNR) \cite{si-snr} of the validation dataset did not increase after 5 consecutive epochs, and gradient norm clipping was set to 0.5 for 100 epochs. The model parameters were the same as a base model in \cite{mcss8} except for the number of output channels of the decoder ($2 \rightarrow 2M$). The loss function for training was the summation of phase-constrained magnitude (PCM) losses \cite{pcm} calculated for individual channels of the multichannel output.



\begin{figure}[htbp]
\centerline{\includegraphics[width=\columnwidth]{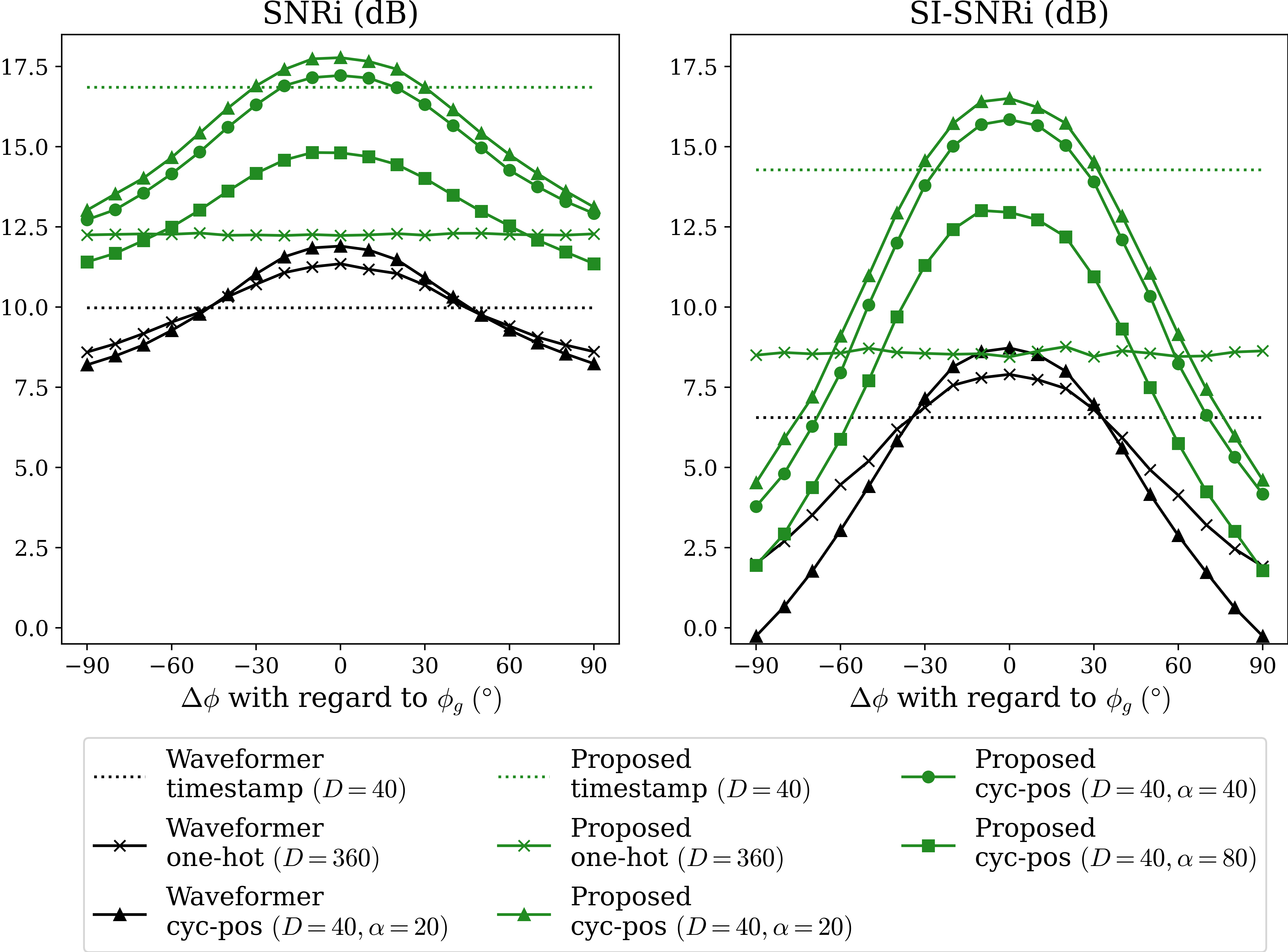}}
\caption{SNRi and SI-SNRi change with respect to target azimuth angle.}
\label{fig_sensitivity}
\end{figure}

\begin{figure}[htbp]
\centerline{\includegraphics[width=\columnwidth]{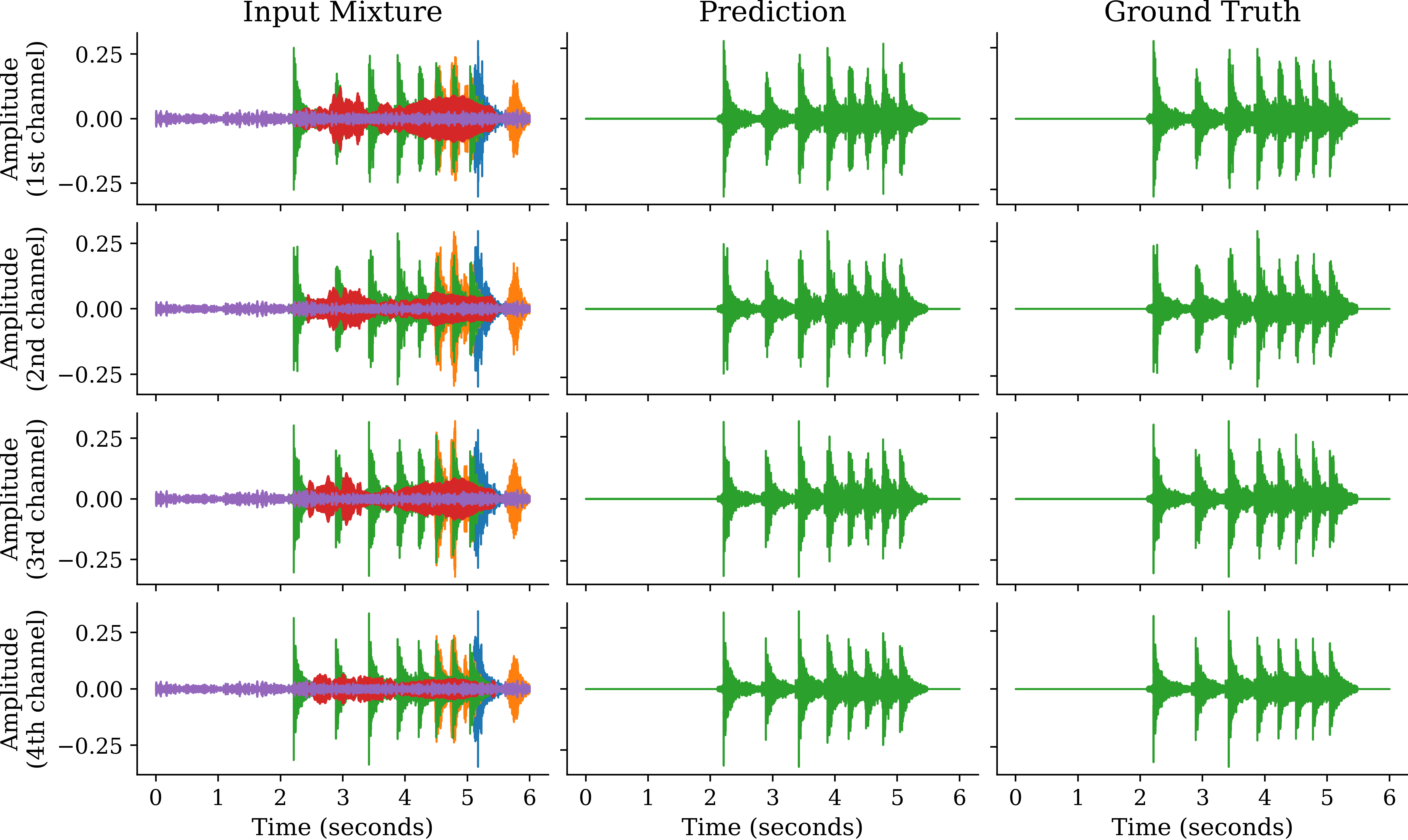}}
\caption{Multichannel waveforms of input mixture, extracted from the model, and ground truth.}
\label{fig_waveform}
\end{figure}

\subsection{Analysis of Results}
We evaluated the effectiveness of our method by measuring the improvement in SNR (SNRi) and SI-SNR (SI-SNRi) compared to the input mixture. These metrics assess the models' ability to suppress other sounds and noise while preserving the target signal. To assess how well the interchannel relations between microphone pairs were maintained, we also calculated mean-absolute-error (MAE) of interchannel metrics (Spatial Errors), such as $\Delta$ILD, $\Delta$IPD, $\Delta$ITD, and $\Delta$ITD-GCC\footnote{Average of the absolute differences between the model prediction and ground truth in interchannel level, phase, time differences using simple cross-correlation, and ITD using generalized cross-correlation phase transform (GCC-PHAT) calculated on all pairs, respectively}, utilized in previous studies \cite{semantic-hearing, interaural}. The comparative results are shown in Table \ref{tab_result}.

As a baseline model for comparison, we modified the binaural extraction model\footnote{\url{https://github.com/vb000/SemanticHearing}}~\cite{semantic-hearing} based on Waveformer \cite{real-time} into a 4-channel model. Compared to the baseline, our model performs significantly better, yielding higher SNRs and lower spatial errors. This result emphasizes the importance of encoding spatio-temporal information using a complex spectrogram instead of a waveform directly for extracting a multichannel sound. Additionally, the extraction performance is noticeably better when using a cyc-pos vector rather than a one-hot-encoded vector. Since the cyc-pos vector has smooth variation across azimuthal angles and has periodicity unlike the one-hot vector, it better integrates into spatially encoded features. However, with large values of $\alpha$ especially greater than the embedding dimension, performance is degraded because the embedding is no longer smooth and changes rapidly in a short cycle even for closely located directions. Meanwhile, most models employ similar $\Delta$IPD values, encountering difficulties in analyzing interchannel phase differences directly, rather than the level or time difference of arrival. 

To evaluate the sensitivity to incorrect directional clues, we measured extraction performance across gradually changing directional clues with a resolution of $10^\circ$ from the true target direction. Results presented in Fig.~\ref{fig_sensitivity} indicate that cyc-pos vectors maintain robust extraction performance near the true target direction, showing less than $1$\,dB decrease in SNRi for azimuth angle difference of $\pm 20^\circ$. 
In contrast, the one-hot vector embedding showed no variation with changing azimuth angles, indicating that the spatial information was not utilized by the model. 
Thus, using cyc-pos vectors is more advantageous, as it not only reduces memory usage by lowering the embedding dimension but also maintains periodicity. 

One example of M2M-TSE is presented in Fig.~\ref{fig_waveform}, for the cyc-pos vector with parameters $D=40$ and $\alpha=20$. The 4-channel waveforms extracted by the proposed model demonstrate that the target source signals are well extracted despite significant overlaps with other sound sources in time. 
Demo is available at \url{https://choishio.github.io/demo_M2M-TSE/}.

\section{Conclusion}
We introduced an M2M-TSE framework for extracting multichannel sound from diverse sound mixtures using target direction and timestamp clues. The direction clue, embedded in form of cyclic positional encoding, was directly integrated with multichannel features of modified DeFTAN-II blocks to enable multichannel sound extraction. The proposed model demonstrated superior performance across multiple channels, outperforming the state-of-the-art extraction model using the same types of clues. This approach to extracting multichannel signals paves the way for separating and editing multichannel audio recordings by preserving spatial information of individual sources. 

\vfill\pagebreak
\bibliographystyle{IEEEbib}
\bibliography{references}

\begin{thebibliography}{10}

\bibitem{cocktail}
S.~Haykin and Z.~Chen,
\newblock ``The cocktail party problem,''
\newblock {\em Neural computation}, vol. 17, no. 9, pp. 1875--1902, 2005.

\bibitem{uss}
T.~Ochiai, M.~Delcroix, Y.~Koizumi, H.~Ito, K.~Kinoshita, and S.~Araki,
\newblock ``Listen to what you want: Neural network-based universal sound selector,''
\newblock in {\em Proc. Interspeech}, Shanghai, China, 2020, ISCA, pp. 1441--1445.

\bibitem{real-time}
B.~Veluri, J.~Chan, M.~Itani, T.~Chen, T.~Yoshioka, and S.~Gollakota,
\newblock ``Real-time target sound extraction,''
\newblock in {\em Proc. IEEE Int. Conf. Acoust., Speech, Signal Process. (ICASSP)}, Rhodes, Greece, 2023, IEEE, pp. 1--5.

\bibitem{soundbeam}
M.~Delcroix, J.~B. V{\'a}zquez, T.~Ochiai, K.~Kinoshita, Y.~Ohishi, and S.~Araki,
\newblock ``Soundbeam: Target sound extraction conditioned on sound-class labels and enrollment clues for increased performance and continuous learning,''
\newblock {\em IEEE/ACM Trans. Audio, Speech, Lang. Process. (TASLP)}, vol. 31, pp. 121--136, 2022.

\bibitem{conceptbeam}
Y.~Ohishi, M.~Delcroix, T.~Ochiai, S.~Araki, D.~Takeuchi, K.~Kashino, et~al.,
\newblock ``Conceptbeam: Concept driven target speech extraction,''
\newblock in {\em Proc. 30th ACM Int. Conf. Multimedia}, Lisboa, Portugal, 2022, pp. 4252--4260.

\bibitem{timestamp1}
H.~Wang, D.~Yang, C.~Weng, J.~Yu, and Y.~Zou,
\newblock ``Improving target sound extraction with timestamp information,''
\newblock in {\em Proc. Interspeech}, Incheon, Korea, 2022, ISCA, pp. 1526--1530.

\bibitem{timestamp2}
D.~Kim, M.~S. Baek, Y.~Kim, and J.~H. Chang,
\newblock ``Improving target sound extraction with timestamp knowledge distillation,''
\newblock in {\em Proc. IEEE Int. Conf. Acoust., Speech, Signal Process. (ICASSP)}, Seoul, Korea, 2024, IEEE, pp. 1396--1400.

\bibitem{describe}
X.~Liu, Q.~Kong, Y.~Zhao, H.~Liu, Y.~Yuan, W.~Wang, et~al.,
\newblock ``Separate anything you describe,''
\newblock {\em arXiv preprint arXiv:2308.05037}, 2023,
\newblock [Online]. Available: https://arxiv.org/abs/2308.05037.

\bibitem{describe2}
X.~Liu, H.~Liu, Q.~Kong, X.~Mei, J.~Zhao, W.~Wang, et~al.,
\newblock ``Separate what you describe: Language-queried audio source separation,''
\newblock in {\em Proc. Interspeech}, Incheon, Korea, 2022, ISCA, pp. 1801--1805.

\bibitem{rezero}
R.~Gu and Y.~Luo,
\newblock ``Rezero: Region-customizable sound extraction,''
\newblock {\em IEEE/ACM Trans. Audio, Speech, Lang. Process. (TASLP)}, vol. 32, pp. 2576--2589, 2024.

\bibitem{multimodal}
C.~Li, Y.~Qian, Z.~Chen, D.~Wang, T.~Yoshioka, M.~Zeng, et~al.,
\newblock ``Target sound extraction with variable cross-modality clues,''
\newblock in {\em Proc. IEEE Int. Conf. Acoust., Speech, Signal Process. (ICASSP)}, Rhodes, Greece, 2023, IEEE, pp. 1--5.

\bibitem{clapsep}
H.~Ma, Z.~Peng, M.~Shao, J.~Liu, X.~Li, and X.~Wu,
\newblock ``Clapsep: Leveraging contrastive pre-trained models for multi-modal query-conditioned target sound extraction,''
\newblock {\em arXiv preprint arXiv:2402.17455}, 2024,
\newblock [Online]. Available: https://arxiv.org/abs/2402.17455.

\bibitem{hetero}
E.~Tzinis, G.~Wichern, A.~Subramanian, P.~Smaragdis, and J.~L. Roux,
\newblock ``Heterogeneous target speech separation,''
\newblock in {\em Proc. Interspeech}, Incheon, Korea, 2022, ISCA, pp. 1796--1800.

\bibitem{oct}
E.~Tzinis, G.~Wichern, P.~Smaragdis, and J.~L. Roux,
\newblock ``Optimal condition training for target source separation,''
\newblock in {\em Proc. IEEE Int. Conf. Acoust., Speech, Signal Process. (ICASSP)}, Rhodes, Greece, 2023, IEEE, pp. 1--5.

\bibitem{mcss1}
R.~Gu, J.~Wu, S.~X. Zhang, L.~Chen, Y.~Xu, D.~Yu, et~al.,
\newblock ``End-to-end multi-channel speech separation,''
\newblock {\em arXiv preprint arXiv:1905.06286}, 2019,
\newblock [Online]. Available: https://arxiv.org/abs/1905.06286.

\bibitem{mcss2}
J.~Zhang, C.~Zoril{\u{a}}, R.~Doddipatla, and J.~Barker,
\newblock ``On end-to-end multi-channel time domain speech separation in reverberant environments,''
\newblock in {\em Proc. IEEE Int. Conf. Acoust., Speech, Signal Process. (ICASSP)}, Virtual, 2020, IEEE, pp. 6389--6393.

\bibitem{mcss3}
R.~Gu, S.~X. Zhang, L.~Chen, Y.~Xu, M.~Yu, D.~Su, Y.~Zou, and D~Yu,
\newblock ``Enhancing end-to-end multi-channel speech separation via spatial feature learning,''
\newblock in {\em Proc. IEEE Int. Conf. Acoust., Speech, Signal Process. (ICASSP)}, Virtual, 2020, IEEE, pp. 7319--7323.

\bibitem{mcss4}
K.~Tesch and T.~Gerkmann,
\newblock ``Multi-channel speech separation using spatially selective deep non-linear filters,''
\newblock {\em IEEE/ACM Trans. Audio, Speech, Lang. Process. (TASLP)}, vol. 32, pp. 542--553, 2023.

\bibitem{mcss5}
D.~Lee and J.~W. Choi,
\newblock ``Deft-an: Dense frequency-time attentive network for multichannel speech enhancement,''
\newblock {\em IEEE Signal Process. Letters (SPL)}, vol. 30, pp. 155--159, 2023.

\bibitem{mcss6}
S.~Wang, X.~Kong, X.~Peng, H.~Movassagh, V.~Prakash, and Y.~Lu,
\newblock ``Dasformer: Deep alternating spectrogram transformer for multi/single-channel speech separation,''
\newblock in {\em Proc. IEEE Int. Conf. Acoust., Speech, Signal Process. (ICASSP)}, Rhodes Island, Greece, 2023, IEEE, pp. 1--5.

\bibitem{mcss7}
Z.~Q. Wang, S.~Cornell, S.~Choi, Y.~Lee, B.~Y. Kim, and S.~Watanabe,
\newblock ``Tf-gridnet: Integrating full-and sub-band modeling for speech separation,''
\newblock {\em IEEE/ACM Trans. Audio, Speech, Lang. Process. (TASLP)}, vol. 31, pp. 3221--3236, 2023.

\bibitem{mcss8}
D.~Lee and J.~W. Choi,
\newblock ``Deftan-ii: Efficient multichannel speech enhancement with subgroup processing,''
\newblock {\em arXiv preprint arXiv:2308.15777}, 2023,
\newblock [Online]. Available: https://arxiv.org/abs/2308.15777.

\bibitem{mcdoa1}
M.~A. Chung, C.~W. Lin, and H.~C. Chou,
\newblock ``Combined multi-sensor based angle clipping algorithm and multi-channel noise removal method for multi-channel sound localization,''
\newblock {\em IEEE Sens. J.}, vol. 24, no. 1, pp. 700--709, 2023.

\bibitem{mcdoa2}
H.~Taherian, A.~Pandey, D.~Wong, B.~Xu, and D.~Wang,
\newblock ``Leveraging sound localization to improve continuous speaker separation,''
\newblock in {\em Proc. IEEE Int. Conf. Acoust., Speech, Signal Process. (ICASSP)}, Seoul, Korea, 2024, IEEE, pp. 621--625.

\bibitem{dse}
A.~Pandey, S.~Lee, J.~Azcarreta, D.~Wong, and B.~Xu,
\newblock ``All neural low-latency directional speech extraction,''
\newblock in {\em Proc. Interspeech}, Kos, Greece, 2024, ISCA, pp. 4328--4332.

\bibitem{semantic-hearing}
B.~Veluri, M.~Itani, J.~Chan, T.~Yoshioka, and S.~Gollakota,
\newblock ``Semantic hearing: Programming acoustic scenes with binaural hearables,''
\newblock in {\em Proc. 36th Annual ACM Symp. User Interface Software Tech. (UIST)}, San Francisco, CA, USA, 2023, pp. 1--15.

\bibitem{binaural-se}
V.~Tokala, E.~Grinstein, M.~Brookes, S.~Doclo, J.~Jensen, and P.~A. Naylor,
\newblock ``Binaural speech enhancement using deep complex convolutional transformer networks,''
\newblock in {\em Proc. IEEE Int. Conf. Acoust., Speech, Signal Process. (ICASSP)}, Seoul, Korea, 2024, IEEE, pp. 681--685.

\bibitem{binaural-filtering}
T.~J. Klasen, S.~Doclo, T.~Van~den Bogaert, M.~Moonen, and J.~Wouters,
\newblock ``Binaural multi-channel wiener filtering for hearing aids: preserving interaural time and level differences,''
\newblock in {\em Proc. IEEE Int. Conf. Acoust., Speech, Signal Process. (ICASSP)}, Toulouse, France, 2006, IEEE, vol.~5, pp. V--V.

\bibitem{interaural}
C.~Hernandez-Olivan, M.~Delcroix, T.~Ochiai, N.~Tawara, T.~Nakatani, and S.~Araki,
\newblock ``Interaural time difference loss for binaural target sound extraction,''
\newblock in {\em Proc. Int. Workshop Acoust. Signal Enhancement (IWAENC)}, Aalborg, Denmark, 2024, IEEE, pp. 210--214.

\bibitem{dense}
G.~Huang, Z.~Liu, L.~Van Der~Maaten, and K.~Q. Weinberger,
\newblock ``Densely connected convolutional networks,''
\newblock in {\em Proc. IEEE Conf. Comp. Vis. Pattern Recognit. (CVPR)}, Honolulu, HI, 2017, pp. 4700--4708.

\bibitem{cyc-pos}
H.~Lee, C.~Homeyer, R.~Herzog, J.~Rexilius, and C.~Rother,
\newblock ``Spatio-temporal outdoor lighting aggregation on image sequences using transformer networks,''
\newblock {\em Int. J. Comp. Vis. (IJCV)}, vol. 131, no. 4, pp. 1060--1072, 2023.

\bibitem{layernorm}
J.~Lei~Ba, J.~R. Kiros, and G.~E. Hinton,
\newblock ``Layer normalization,''
\newblock {\em arXiv preprint arXiv:1607.06450}, 2016,
\newblock [Online]. Available: https://arxiv.org/abs/1607.06450.

\bibitem{prelu}
K.~He, X.~Zhang, S.~Ren, and J.~Sun,
\newblock ``Delving deep into rectifiers: Surpassing human-level performance on imagenet classification,''
\newblock in {\em Proc. IEEE Int. Conf. Comp. Vis. (ICCV)}, Santiago, Chile, 2015, pp. 1026--1034.

\bibitem{fsd-kaggle}
E.~Fonseca, M.~Plakal, F.~Font, D.~P.~W. Ellis, X.~Favory, X.~Serra, et~al.,
\newblock ``General-purpose tagging of freesound audio with audioset labels: Task description, dataset, and baseline,''
\newblock in {\em Proc. Workshop Detect. Classif. Acoust. Scenes Events (DCASE)}, Surrey, UK, 2018, pp. 69--73.

\bibitem{spatialize}
Z.~Q. Wang and D.~L. Wang,
\newblock ``Multi-microphone complex spectral mapping for speech dereverberation,''
\newblock in {\em Proc. IEEE Int. Conf. Acoust., Speech, Signal Process. (ICASSP)}, Virtual, 2020, IEEE, pp. 486--490.

\bibitem{pyroom}
R.~Scheibler, E.~Bezzam, and I.~Dokmani{\'c},
\newblock ``Pyroomacoustics: A python package for audio room simulation and array processing algorithms,''
\newblock in {\em Proc. IEEE Int. Conf. Acoust., Speech, Signal Process. (ICASSP)}, Calgary, Alberta, Canada, 2018, IEEE, pp. 351--355.

\bibitem{scaper}
J.~Salamon, D.~MacConnell, M.~Cartwright, P.~Li, and J.~P. Bello,
\newblock ``Scaper: A library for soundscape synthesis and augmentation,''
\newblock in {\em IEEE Workshop Appl. Signal Process. Audio Acoust. (WASPAA)}, New Paltz, NY, USA, 2017, IEEE, pp. 344--348.

\bibitem{reverb}
K.~Kinoshita, M.~Delcroix, T.~Yoshioka, T.~Nakatani, E.~Habets, R.~Maas, et~al.,
\newblock ``The reverb challenge: A common evaluation framework for dereverberation and recognition of reverberant speech,''
\newblock in {\em IEEE Workshop Appl. Signal Process. Audio Acoust. (WASPAA)}, New Paltz, NY, USA, 2013, IEEE, pp. 1--4.

\bibitem{si-snr}
J.~Le~Roux, S.~Wisdom, H.~Erdogan, and J.~R. Hershey,
\newblock ``Sdr--half-baked or well done?,''
\newblock in {\em Proc. IEEE Int. Conf. Acoust., Speech, Signal Process. (ICASSP)}, Brighton, UK, 2019, IEEE, pp. 626--630.

\bibitem{pcm}
A.~Pandey and D.~L. Wang,
\newblock ``Dense cnn with self-attention for time-domain speech enhancement,''
\newblock {\em IEEE/ACM Trans. Audio, Speech, Lang. Process. (TASLP)}, vol. 29, pp. 1270--1279, 2021.

\end{thebibliography}

\end{document}